# LGBTQ Privacy Concerns on Social Media

**Christine Geeng**

University of Washington

Computer Science and Engineering

Seattle, Washington USA

cgeeng@cs.washington.edu

**Alexis Hiniker**

University of Washington

Information School

Seattle, Washington USA

alexisr@uw.edu

## Abstract
We conducted semi-structured interviews with members of the LGBTQ community about their privacy practices and concerns on social networking sites. Participants used different social media sites for different needs and adapted to not being completely out on each site. We would value the opportunity to discuss the unique privacy and security needs of this population with workshop participants and learn more about the privacy needs of other marginalized user groups from researchers who have worked in those communities.

## Author Keywords
LGBTQ users; social computing; privacy.

## ACM Classification Keywords
H.5.m. Information interfaces and presentation (e.g., HCI): Miscellaneous

## Introduction
Online networks provide LGBTQ+ individuals with "safe and accepting environments online" to "address challenges they may face in their daily lives, such as social isolation" [1, 7], but social networking sites (SNS) can also be risky for identity disclosure due to audience context collapse and persistent digital traces [1].

Prior work suggests that designers currently address privacy management issues by focusing on the needs of a population using a single SNS [13]. However, people increasingly use multiple SNSs [3], and in particular, queer individuals often maintain multiple profiles for different audiences [1, 6]. We ask: **what are the privacy concerns and practices of LGBTQ individuals that span** **their use of multiple SNSs?** Our goal in exploring this question is to inform the design of SNSs by providing a higher-fidelity representation of the needs and practices of this user group.

## Background, Related Work, and Methods

Many researchers have explored how theories of boundary regulation and disclosure practices surface in social media usage [2, 8, 9, 13, 15], particularly for Facebook. Fox and colleagues looked at how LGBT individuals, at various levels of outness, communicate about identity and LGBT+ issues on Facebook through a co-cultural theory lens [6].

They, along with Blackwell et al. note that several participants use a secondary SNS for queer self-expression, which leads us to ask how people manage privacy for multiple SNSs [1]. To better understand these practices, we conducted semi-structured interviews with LGBTQ+ individuals who use multiple social media sites (including Facebook, Twitter, and Instagram) and asked how they manage their identity across sites.

## Results: Privacy Concerns

RISK MODELING
Interviewees discussed who they did *not* want to learn about their queer identity, how likely it was that their queer identity would be discovered, and how this affected their behavior when interacting with others or sharing queer content on their pages. P4 and P6 decided that the risk of discovery and potential impact were not enough to warrant disclosing less information (see sidebar for examples).

ACCOUNTING FOR OTHERS
P10 and P2 mentioned being cognizant of their partner's level of outness (see sidebar). On Instagram, P9 asks his boyfriend not to tag him in his photos because P9 does not want his brother and cousins, who follow him on Instagram, to potentially see that content.

PLAUSIBLE DENIABILITY
P4, P5, and P6 mentioned being comfortable interacting with visibly queer content because it could be interpreted as being an ally and left plausible deniability of identifying with that gender or orientation. P6, who is visibly bisexual on Facebook but not visibly transgender, mentioned checking "Going To" on trans Facebook events because, "*[My partner] and I are both trans, and I tell people that he's trans even if I don't tell them [I am]. They're probably just thinking I'm a good ally.*"

## Results: Privacy Strategies

*Using Existing Tools*
*Blocking*: P5's Instagram audience is composed only of people who know she is transgender, so regarding others: "*I did find my brother on there, so I blocked his account to be safe. Cause I want to come out to him on my own terms.*" P1 and P5 also use blocking on Instagram and on Snapchat.

*Partitioning Social Media Space:* P11 and P4 used private groups on Facebook to make connections with other queer women. P4 will not click "Interested" or "Going" on a public LGBTQ event, because she has family she is not out to on Facebook. But she will select one of these options if the event is private or posted in a closed group, because that information will not show up in her public feed. Before P9 was generally out, he created a privacy group with whom he could safely share queer content on Facebook. As he came out to more people and the list for sharing queer content grew bigger, he switched to using a privacy group containing people he is not out to. He then made public posts that weren't visible to this second group.

*Adapting to Lack of Tools*
*Obfuscation:* P9 noted that events and likes are publicly visible, and *"I'm pretty sure [my mom] has a good chance of seeing on her scrolling sidebar."* His response is, *"I don't really like a lot of pages…and I've done those at really odd hours in the night. Um, same thing with event invites. I, if I do say I'm going to something that's remotely gay, I will try to immediately put interested on six or seven other events, so it looks like I just mass-okayed a bunch of friend requests for events."*

*Self-Censorship:* P4 and P9 say that at times they choose not to perform an action they would otherwise enact, because it was related to queerness that could be visible (see sidebar).

*Shifting Sites:* P5 moved from Facebook to Instagram where she felt could be completely out. P9 and P1 did not allow Instagram to import contacts from their phone or Facebook to suggest people to follow because they only wanted an audience of close friends so they wouldn't have to "curate a lot of the posts" (P1). P10 stopped using Facebook and uses Snapchat, where she has complete control over data flow, to only communicate with people she is out to and close with.

### Results: Non-Privacy Motivations
P1 and P9 used blocking but were conflicted with maintaining socially appropriate behavior. P1 blocked an acquaintance from high school on Instagram, but later unblocked her after learning the "unspoken rules of Instagram," where blocking was a serious violation of social norms and could be perceived as rude. P9 expressed similar understanding of blocking propriety; when using Snapchat at a Pride Parade, he blocked his family members who he was not out to, and then unblocked them the next day. This served as a compromise between adhering to Snapchat etiquette and protecting his privacy.

### DISCUSSION
These experiences stress how important it is to reconsider a "one-size fits all" approach to privacy. While some findings can be abstracted to support existing privacy theory, such as the importance of temporality in boundary regulation [12] and managing group co-presence [8], these interviews call attention to the high risk of being accidentally outed specific to the LGBTQ community and the precarious ways identity information can be revealed.

Blackwell and colleagues note that, "in the context of shifting social movements wherein public opinion is constantly evolving, an individual's privacy is heavily dependent on…the temporal context in which it is instantiated." While 92% of LGBTQ+ Americans say that society has become more accepting, about 4-in-10 report "being rejected by a family member or close friend because of their orientation" [14]. P6 stated that he is out as bi but not as transgender on Facebook because he feels that LGB individuals are more accepted and face less stigma than transgender people. Until there is more widespread acceptance of minorities of both sexual orientation and gender identity, these privacy concerns remain relevant.

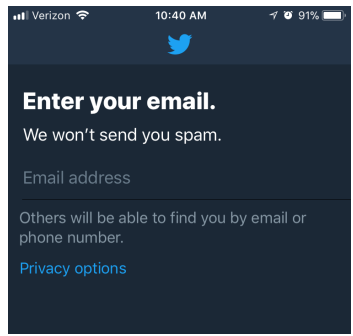

TWITTER DISCOVERABILITY
Discoverability by email or phone are on by default when a user signs up.

The user must click through to Privacy Options and manually turn off these features.

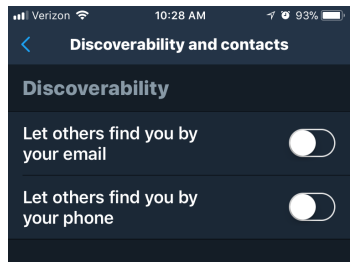

Expanding on Omarzu's disclosure decision model [11] and Vitak and Kim's extension [15], we found that disclosure goals also include solidarity and advocacy, consistent with Blackwell et al.'s findings of LGBT parents on Facebook being incidental advocates. P1, P2, P4, P5, P6, P8, P9, and P11 all stated they felt happy when they saw a friend expressing their queer identity on a SNS. Several participants said they posted news about queer issues to advocate for political change. Others expressed joy when their own posts on queerness generated positive support and solidarity from their audience. These are important goals for queer individuals and for LGBTQ acceptance in the US, which are often weighed against the need for selective privacy (as well as other motivations). We make some design suggestions to support successful privacy management and safer disclosure decisions.

*Design Considerations*
Posting is not the only type of data leakage that can pose risks for queer identity. A users' social network may be cognizant of what events the user likes or is interested in, what posts they like, their friends list, and other people's photos they are in. Facebook could provide an option to hide all public likes from a Friends list.

Instagram and Twitter have a feature where users can discover contacts from their phonebook or Facebook. This became inconvenient for several people who expressly created a different SNS account to have a more curated audience; this shows that not even moving to another site is a foolproof boundary regulation. This feature could be off by default.

For participants using SNSs with their public/real name, they sometimes had to contend with people they knew in person adding them that they weren't out to. P1, P5, and P9 discussed blocking people on Instagram and Snapchat, with P9 having to navigate the tension of potentially blocking family members. His ad hoc solution was temporarily blocking them when he was at a queer event. These limited privacy SNSs could benefit from soft temporary blocking, where people who are blocked are not informed and are only blocked for a set duration.

These findings demonstrate systematic ways in which SNSs both support members of the LGBTQ community and make them more vulnerable. We hope to share these population-specific privacy considerations with members of the workshop and to discuss ways in which our findings might translate to other groups. By exploring how this population navigates online spaces, we not only illuminate design considerations specific to LGBTQ individuals, we also provide a case study in the importance of differentiated privacy design.

## Conclusion
This study provides a more holistic view of LGBTQ use of social media sites, and how individuals balance certain needs while keeping their queer identity private from people in their social networks that they are not out to. As new SNS features for sharing information come out, we should be cognizant of how they may affect marginalized populations. This calls into question how default connectivity in SNSs should be weighed against the safety of marginalized users. We look forward to sharing these themes at the workshop, hearing from others designing for other populations, and exploring how findings about the privacy needs of other populations might shape our next steps in this research investigation.

**Acknowledgements**

We thank the participants and colleagues who provided helpful comments on previous versions of this document. We gratefully acknowledge the grant from NSF (CNS-1463968).

*Computing - CSCW '14*, 461–474.
https://doi.org/10.1145/2531602.2531672

**Appendix**

*Methods*

I conducted semi-structured interviews with 11 queer[1] individuals on their use of mainstream social media sites (including Facebook, Twitter, Snapchat, Reddit, and Tumblr).

I posted recruitment messages on Facebook and Twitter, as well as recruited through word of mouth, using my personal network. Through a screening survey, I selected individuals who use at least two SNSs and use at least one daily; I also selected a mix of individuals living in rural, urban, or suburban environments. Participants received a $10 Amazon Gift Card. The highest age was 33 years old, and only people of Caucasian or Asian descent were sampled.

When interviewing, I asked participants to identify their gender identity and sexual orientation, as well as who they are out and not out to. Some people were completely out in terms of their sexual orientation, but partially out for the gender identity. To allow participants to frame their own social media experiences, rather than prompting them with thoughts about security and privacy, I initially asked if identity-related incidents occurred that made them happy or upset. I followed up with specific questions about privacy practices. After interviewing participants either over Skype or in-person, I transcribed the interviews and took an inductive approach to iteratively code the interview data, using boundary regulation as a foundational framework.

*Participant Demographics*

|    | Age | Gender Identity | Sexual Orientation | Region |
|----|-----|-----------------|--------------------|--------|
| P1 | 19  | Non-binary      | Lesbian/Gay        | Massachusetts |
| P2 | 24  | Cis woman       | Prefer women       | Washington |
| P3 | 28  | Cis woman       | Lesbian            | Washington |
| P4 | 33  | Cis Woman       | Lesbian            | Illinois |
| P5 | 26  | Trans woman     | Lesbian            | New York |
| P6 | 26  | Trans male      | Bisexual           | Indiana |
| P7 | 25  | Cis male        | Gay                | Washington |

---

[1] For this paper, the term "LGBTQ+," "queer," or "sexual minority" is used to refer to the population of interest, which includes anyone who does not identify as exclusively heterosexual (including gay, lesbian, bisexual, pansexual, asexual, transgender, non-binary, and others).

| P8  | 23 | Gender fluid | Queer, Asexual | California |
| P9  | 23 | Cis male   | Gay       | Washington |
| P10 | 24 | Cis woman  | Bisexual  | Washington |
| P11 | 26 | Cis woman  | Lesbian   | Illinois   |